# Capabilities and prospects of the East Asia Very Long Baseline Interferometry Network

T. An[1,2,*], B. W. Sohn[3], H. Imai[4]

1. Shanghai Astronomical Observatory, Key Laboratory of Radio Astronomy, Chinese Academy of Sciences, Nandan Road 80, Shanghai 200030, China. *e-mail: antao@shao.ac.cn

2. Key Laboratory of Cognitive Radio and Information Processing, Guilin University of Electronic Technology, 541004 Guilin, China. 3Korea Astronomy and

3. Space Science Institute, 776 Daedeokdae-ro, Yuseong-gu, Daejeon 34055, Republic of Korea.

4. Graduate School of Science and Engineering, Kagoshima University, 1-21-35 Korimoto, Kagoshima 890-0065, Japan.

## Abstract

The very long baseline interferometry (VLBI) technique offers angular resolutions superior to any other instruments at other wavelengths, enabling unique science applications of high-resolution imaging of radio sources and high-precision astrometry. The East Asia VLBI Network (EAVN) is a collaborative effort in the East Asian region. The EAVN currently consists of 21 telescopes with diverse equipment configurations and frequency setups, allowing flexible subarrays for specific science projects.  The EAVN provides the highest resolution of 0.5 mas at 22 GHz, allowing the fine imaging of jets in active galactic nuclei, high-accuracy astrometry of masers and pulsars, and precise spacecraft positioning. The soon-to-be-operational Five-hundredmeter Aperture Spherical radio Telescope (FAST) will open a new era for the EAVN. This state-of-the-art VLBI array also provides easy access to and crucial training for the burgeoning Asian astronomical community. This Perspective summarizes the status, capabilities and prospects of the EAVN.

The very long baseline interferometry (VLBI) technique is a unique development based on radio interferometry that enables radio astronomical observations at high angular resolutions of milli-arcsecond and even at submilliarcsecond scales. The current record is held by the Russian space radio telescope RadioAstron satellite [1] with ground-based antennas, resulting in a highest angular resolution of 8 μas at 22 GHz. Recently, the Event Horizon Telescope (EHT)[2] conducted imaging observations of the supermassive black hole (SMBH) in the centre of the Milky Way, Sagittarius A* (Sgr A*), and another SMBH in Messier 87 (M87) with an angular resolution of 25 μas at 230 GHz, which approaches the size of event horizons of these black holes.

VLBI was developed in the 1960s by the requirement for resolving compact radio sources [3,4], and today continues to incorporate the newest technological developments in related fields, for example, frequency standards of hydrogen masers, microwave receivers, digital filtering and recording, and correlators. Advanced image reconstruction and calibration techniques allow high-fidelity images with a super resolution higher than the standard diffraction limit to be obtained, even with a sparsely distributed VLBI network [5,6].

Established astronomical VLBI networks include the Very Long Baseline Array (VLBA) in the United States, the European VLBI Network (EVN), and the Long Baseline Array (LBA) in Australia and South Africa, and others. In 2003, astronomers first proposed to form a combined East Asia VLBI Network (EAVN) to strengthen regional VLBI collaborations. Table 1 compares the major VLBI networks in the Northern Hemisphere.

| Network Name | | Location | Number of telescopes | Diameters of telescopes (m) | Operational Frequency Bands (GHz) | Baseline range (km) | Performance of 22 GHz band | |
|---|---|---|---|---|---|---|---|---|
| | | | | | | | Resolution (mas) | Imaging Sensitivity (mJy beam$^{-1}$) |
| (1) | | (2) | (3) | (4) | (5) | (6) | (7) | (8) |
| EAVN | | China, Japan, Korea | 21 | 11 – 500 | 2.3, 6.7, 8.4, 22, 43 | 6.1 – 5,078 | 0.5 | 0.076 |
| | CVN | China | 6 | 25 – 500 | 1.6, 2.3, 8.4, 22 | 6.1 – 3,249 | 0.9 (2.3 @ 8.4 GHz) | 12.6 (0.08 @ 8.4 GHz) |
| | JVN, VERA, NRO | Japan | 11 | 11 – 64 | 2.3, 6.7, 8.4, 22, 43 | 84 – 2,270 | 1.2 | 0.15 |
| | KVN, NGII | Korea | 4 | 21 – 22 | 8.4, 22, 43, 86, 129 | 120 – 476 | 5.9 | 0.4 |
| EVN | | Europe, East Asia, South Africa | 27 | 15 – 305 | 1.4/1.6, 5, 6.7 -22 | 198 – 10,160 | 0.26 | 0.04 |
| VLBA | | United States | 10 | 25 | 0.3-90 | 236 – 8,611 | 0.32 | 0.1 |

**Table 1: Description of major VLBI networks in the Northern Hemisphere.** NGII - National Geographic Information Institute of Korea. Note that some individual telescopes of EAVN and EVN have broader frequency coverages. VLBA has 11 discrete frequency bands from 0.3 to 90 GHz. As the CVN is normally operational at a maximum 8.4 GHz, the resolution and sensitivity of CVN at 8.4 GHz are also presented in brackets. The 1σ sensitivity value of the EAVN given in the last column is estimated with the telescopes listed in Fig. 1, assuming a one hour integration time and one Gigabits per second (Gbps) data rate. Future EAVN 22 GHz commissioning experiments will verify the sensitivity, including all available EAVN telescopes.

## Current status and capabilities

The EAVN includes most VLBI facilities and astronomical radio telescopes in East Asia: the Chinese VLBI Network (CVN) [7], the Japanese VLBI Network (JVN) [8], the VLBI Exploration of Radio Astrometry (VERA) telescopes [9], the Korean VLBI Network (KVN) [10], and three correlators in Mizusawa (Japan), Daejeon (Korea), and Shanghai (China). Table 2 summarizes the properties of these subarrays.

The CVN was first proposed in the 1970s [7], and the commissioning of the first telescope Sheshan 25 m began in 1987. At present, the CVN consists of 5 telescopes normally operating at 2.3 and 8.4 GHz and a correlator centre in Shanghai. The CVN was mainly used for space exploration, but has expanded its astrophysical applications in recent years. Some individual telescopes (for example, Sheshan 25 m, Tianma 65 m, Urumqi 26 m) have broader frequency coverage, and also join other international VLBI networks. The Five-hundred-meter Aperture Spherical radio Telescope (FAST) [11] completed its construction in southwest China in 2016 [12], and will join CVN and EAVN observations once its VLBI functionality is available.

The Japanese VLBI facilities are jointly operated by universities and the National Astronomical Observatory of Japan (NAOJ), including the JVN consisting of six radio telescopes, VERA consisting of four telescopes, and 45 m Nobeyama telescope. These flexibly organized telescopes have the advantage of ad hoc observations, which is important for experiments triggered by fast structural changes of jets in X-ray binaries, active galactic nuclei (AGNs), galactic methanol and water masers during their flaring events. Equipped with a unique dual-beam receiving system [13] (22 and 43 GHz), VERA is dedicated for astrometry studies using the fringe phase-referencing technique to measure trigonometric parallaxes and proper motions of stellar sources with maser features [14].

The KVN is a dedicated millimetre-wavelength VLBI facility [15] with first fringes being obtained in 2009. For the purpose of high-sensitivity millimetre-wavelength imaging and simultaneous observations of multiple masers, the Korea Astronomy and Space Science Institute developed and installed a unique multi-frequency receiving optics system [16], allowing for simultaneous observations at 22, 43, 86 and 129 GHz bands. By adopting a special source frequency phase-referencing technique, this system enables accurate VLBI astrometry at frequencies as high as 129 GHz [17]. Recently, a new 22 m radio telescope at Sejong station was inaugurated and often jointly operated with the KVN, allowing for amplitude closure of the KVN.

| Subarray | Telescope Name | Diameter of telescopes (m) | Frequency Bands (GHz) | | | | | | | | |
|---|---|---|---|---|---|---|---|---|---|---|---|
| (1) | (2) | (3) | (4) | | | | | | | | |
| | | | 1.6 | 2.3 | 5 | 6.7 | 8.4 | 22 | 43 | 86 | 129 |
| CVN | FAST | 500 | 0.07 – 3 | | | | | | | | |
| | Kunming | 40 | | ● | | ● | ● | | | | |
| | Miyun | 50 | | ● | | | ● | | | | |
| | Sheshan | 25 | ● | ● | ● | ● | ● | ● | | | |
| | Tianma | 65 | 1.25 – 50 | | | | | | | | |
| | Urumqi | 26 | ● | ● | ● | | ● | ● | | | |
| JVN | Gifu | 11 | | | | | | ● | | | |
| | Hitachi | 32 | | | | ● | ● | ● | | | |
| | Kashima | 34 | ● | ● | | | ● | ● | ● | | |
| | Takahagi | 32 | | | | ● | ● | ● | | | |
| | Usuda | 64 | ● | ● | | | ● | ● | | | |
| | Yamaguchi | 32 | | | | ● | ● | | | | |

| | | | | | | | | | | |
|---|---|---|---|---|---|---|---|---|---|---|
| VERA | Iriki | 20 | | ● | | ● | ● | ● | ● | |
| | Ishigakijima | 20 | | ● | | ● | ● | ● | ● | |
| | Mizusawa | 20 | | ● | | ● | ● | ● | ● | |
| | Ogasawara | 20 | | ● | | ● | ● | ● | ● | |
| NRO | Nobeyama | 45 | | | | | ● | ● | ● | |
| KVN, NGII | Sejong | 22 | | ● | | | ● | ● | ● | |
| | Tamna | 21 | | | | | ● | ● | ● | ● |
| | Ulsan | 21 | | | | 6.4 – 9 | ● | ● | ● | ● |
| | Yonsei | 21 | | | | | ● | ● | ● | ● |

**Table 2 : Parameters of the EAVN telescopes.** The operational frequency bands are marked with solid dots. The available frequency coverage in the early science phase of FAST is 0.07–3 GHz. The frequency coverage of the Tianma telescope is continuous from 1.25 up to 50 GHz. The Ulsan telescope has a broadband receiver at the 6.4-9 GHz frequency range.

The geographical distribution of EAVN telescopes and correlators are presented in Fig. 1. Twenty-one radio telescopes are distributed over 5,000 km from Urumqi in northwestern China to Japan's southeast Ogasawara Islands, and from Mizusawa in northeastern Japan to Kunming in the southwest of China. The EAVN is operational mainly at 6.7, 8.4, 22 and 43 GHz bands with a wider possible capability of observations at 1.6 – 129 GHz. The highest angular resolution at 22 GHz is about 0.5 mas. Owing to the diverse equipment configurations and frequency setups, the EAVN is flexible and able to support specific science projects with its subarrays. Figure 2 shows the typical *u-v* coverage of the EAVN at 22 GHz and comparison with those of the EVN and VLBA. The EAVN comprises densely distributed telescopes in Korea and Japan, thus the *u-v* coverage is richer in short and intermediate baselines. The shortest baselines — a Tianma-Sheshan (~6 km), Kashima-Takahagi/Hitachi (~82 km), and Yensei-Sejong (~120 km) — grant the EAVN an exceptional capability for high-sensitivity imaging of extended structure, similar to inclusion of the e-MERLIN (an array of seven telescopes in UK) in the EVN. The synergy between the EAVN and the EVN+e-MERLIN provides the excellent opportunity for continuous relay monitoring of fast-varying extended sources in the Milky Way.

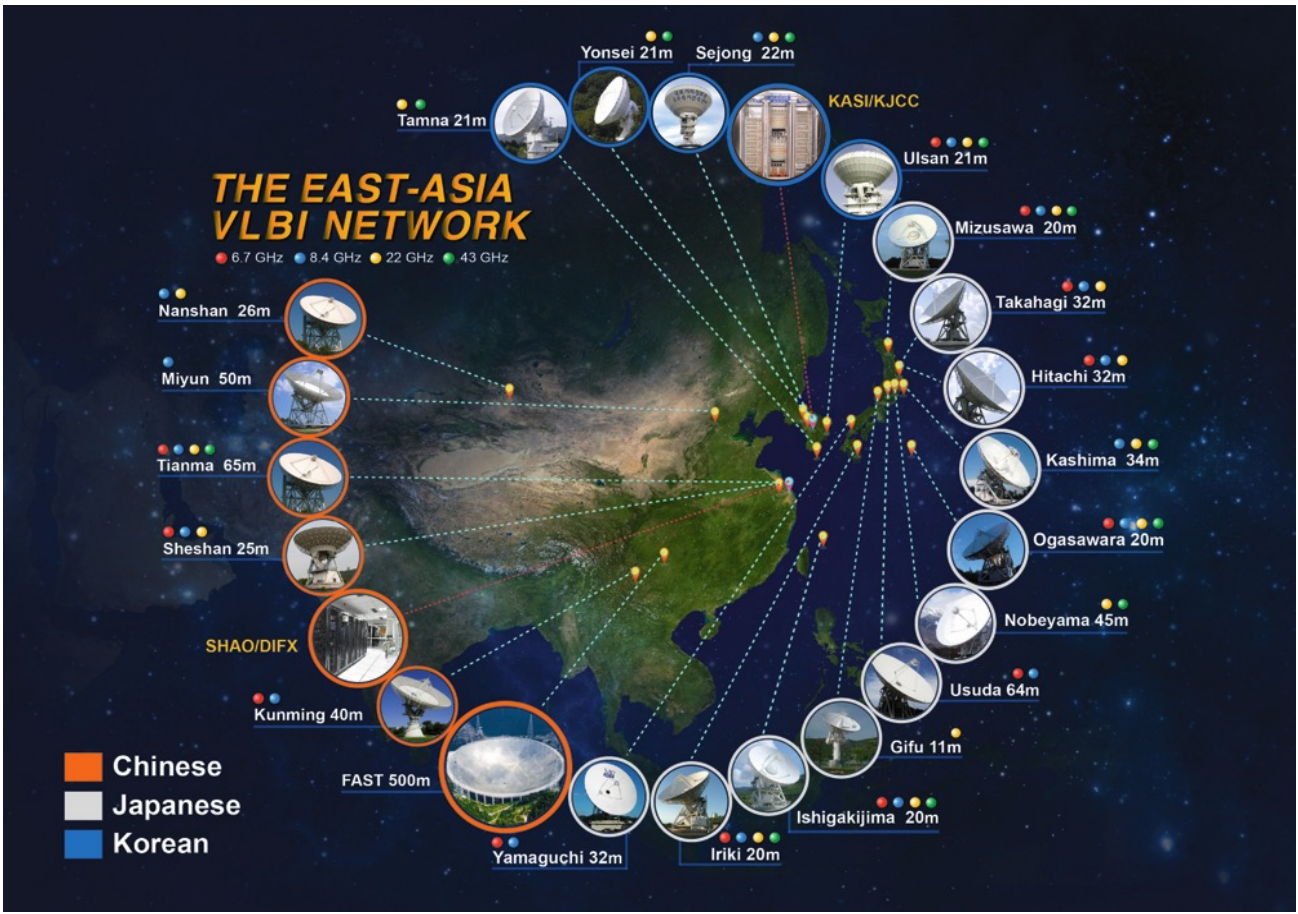

**Fig. 1: Geographic distribution of the EAVN telescopes and correlators.** The EAVN can be flexibly operated using subarrays to support diverse science projects. The coloured circles show telescopes affiliated with each subarray. The small coloured dots above each telescope name indicate their available frequency bands. Note that only four common frequency bands (6.7, 8.4, 22 and 43 GHz) of the EAVN are marked here, but some individual telescopes have broader frequency coverage (see details in Table 2). Credit: Tamna 21 m and Yonsei 21 m, KASI; Sejong 22 m, NGII; KASI/KJCC and Ulsan 21 m, KASI; Mizusawa 20 m, VERA; Takahagi 32 m and Hitachi 32 m, Ibaraki University; Kashima 34 m, NICT; Ogasawara 20 m, VERA; Nobeyama 45 m, Nobeyama Radio Observatory; Usuda 64 m, JAXA; Gifu 11 m, Gifu University; Ishigakijima 20 m and Iriki 20 m, VERA; Yamaguchi 32 m, Yamaguchi University; FAST 500 m, NAOC; Kunming 40 m, SHAO/DIFX, Sheshan 25 m, Tianma 65 m, Miyun 50 m and Nanshan 26 m, CVN; background map, NASA.

The first successful EAVN fringes were obtained with Yonsei (Korea), Mizusawa (Japan) and Sheshan (China) telescopes as part of the KVN commissioning programme in May 2009 [18-19]. Since then, further test observations involving a maximum of 16 EAVN telescopes have been made at 8.4 and 22 GHz to iron out operational, scheduling, correlation, and data reduction issues [20]. Final performance assessment of the network at 6.7, 22 and 43 GHz bands is to be done in 2017, and to be expanded to 2.3 and 8.4 GHz in 2018. Early science operations will commence in early 2018.

The current operation format of the combined KVN and VERA Array (KaVA) will naturally transition to the EAVN once the latter is fully operational, that is, including two calls for proposals, for roughly 500 hours per year. The EAVN management will be governed by the East Asia Core Observatory Association, and the execution of observations will be carried out by dedicated support teams. The KaVA data are now correlated at the Korea–Japan Correlation Center in Daejeon, Korea. In the future, the EAVN data correlation can be shared at two other sites (Shanghai, China and Mizusawa, Japan) depending on the specific science projects and the capability of each individual correlator.

The EAVN is an open VLBI facility for radio astronomers from around the world. Additional telescopes will be inaugurated in the near future, further increasing its capability (see 'Future prospects of the EAVN'). Besides its significant contribution to the global VLBI science, the EAVN provides the burgeoning Asian astronomical community with easy access to a state-of-the-art VLBI array, and plays a crucial role in training new VLBI users. Annual workshops have been organized since 2008, with a focus on exchanging achievements, promoting cooperation and initiating new joint projects. Radio interferometry schools aim to train students and provide hand-on practice sessions for new users.

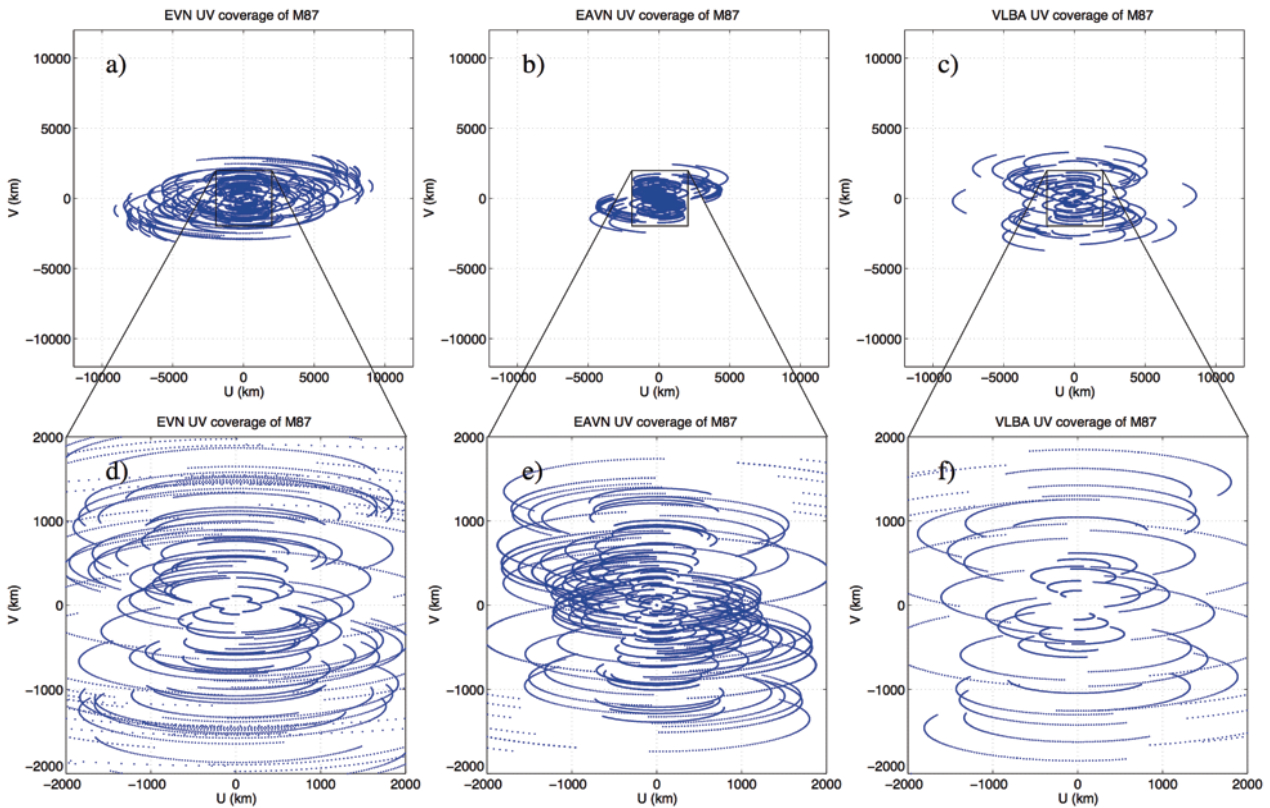

**Fig. 2: Comparison of *u*–*v* coverages on a 12-hour track of the EAVN with EVN and VLBA at 22 GHz frequency band.** The target is M87, a giant radio galaxy lying in the heart of the Virgo cluster. Each dot represents a u–v sampling on a certain baseline in an integration time of five minutes. Telescopes used in creating the u–v coverages: **a,d,** EVN: Effelsberg (Germany), Onsala (Sweden), Torun (Poland), Jodrell Bank, Cambridge (UK), Sardinia, Medicina and Noto (Italy), Metsähovi (Finland), Sheshan (China), Urumqi (China), Yebes (Spain), Svetloe, Zelenchukskaya, Badary (Russia); b,e, EAVN: Gifu, Hitachi, Iriki, Ishigakijima, Kashima, Mizusawa, Nobeyama, Ogasawara, Sejong, Sheshan, Tahahagi, Tamna, Tianma, Ulsan, Urumqi, Yonsei (also seen in Fig. 1); **c,f,** VLBA : Brewster, Fort Davis, Hancock, Kitt Peak, Los Alamos, Mauna Kea, North Liberty, Owens Valley, Pie Town, St. Croix. Panels **d–f** highlight the inner 2,000 km baselines. It is prominent that the EAVN shows a denser sampling in short and intermediate *u–v* spacing. With the availability of a large number of stations, good u–v coverage of the EAVN obtained on long and short baselines enables imaging of both compact and extended celestial structures.

## Science case highlights

As the precursor of the EAVN, KaVA has already been in operation since 2011 observing at 22 and 43 GHz, and has made some progress in studies of AGN jets[21], 44 GHz methanol maser emission from massive star-forming regions[22], and water and silicon-monoxide maser emission in circumstellar envelopes of long-period variable stars[23]. The impact of the short VLBI baselines of KVN on mapping extended structure is well demonstrated in these KaVA results. Since 2016, KaVA started to accept proposals from international users, providing in total 500 hours of open-use time every year.

The key science studies of the EAVN will concentrate on observing high-brightness and compact celestial sources at milliarcsecond and submilliarcseond resolutions (including AGNs, black holes, interstellar and circumstellar masers, and pulsars) and acquiring their precise positions for astrometry and geodesy studies. Figure 3 illustrates the various science projects supported by the EAVN in flexibly selected subarrays and frequency bands. Representative science outcomes obtained with the EAVN telescopes are demonstrated in Figs. 4–6.

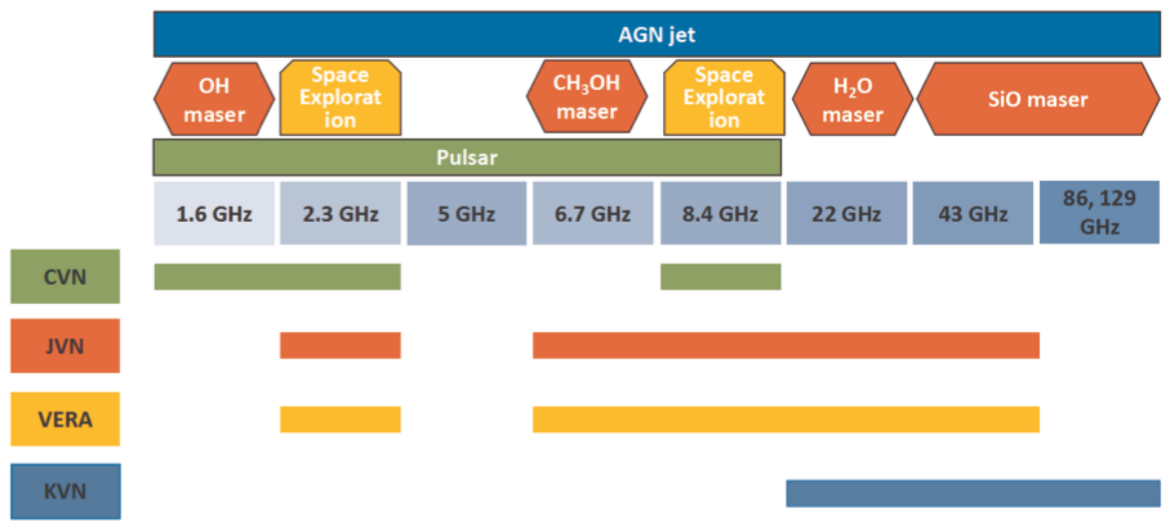

**Fig. 3: A sketch map of the EAVN science cases and corresponding subarrays and frequency setups.** A broad range of operational frequencies and diverse subarray setups facilitate multi-purpose science projects. The inclusion of the FAST 500 m telescope will substantially improve the observing capability of the EAVN at 1.6 and 2.3 GHz. The connection and synergy of the EAVN with other millimetre radio telescope networks (for example, the EVN, VLBA, Atacama Large Millimeter/submillimeter Array and EHT) can significantly contribute to cutting-edge VLBI science of black holes and jets (see discussion in 'Future prospects of the EAVN').

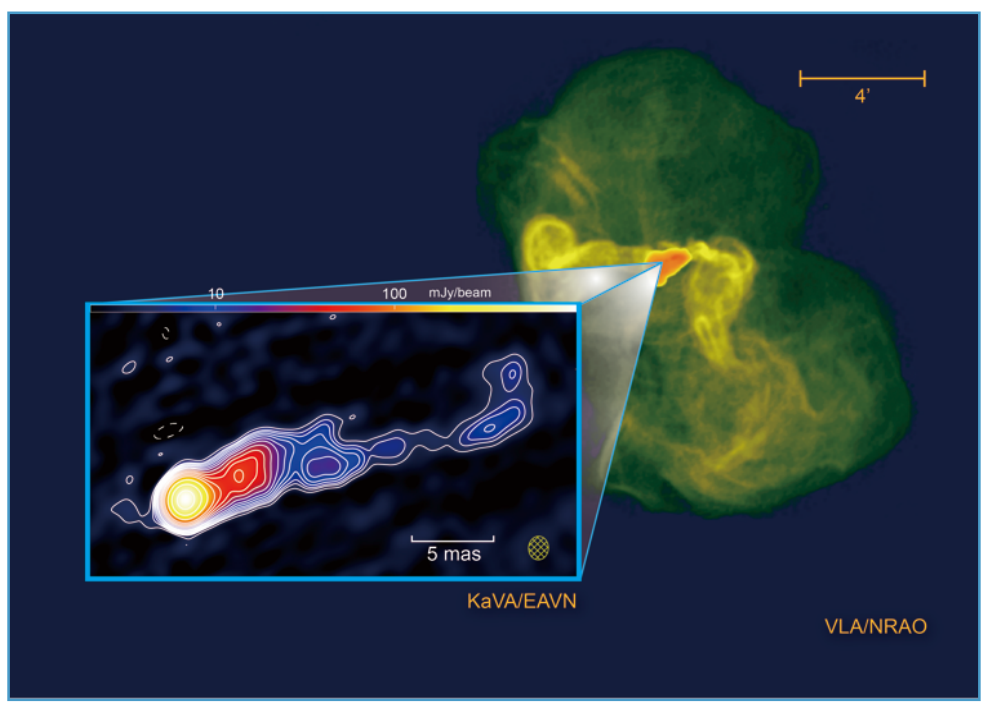

**Fig. 4: The radio structure of the giant radio galaxy M87.** The background image shows emission at 330 MHz imaged with the VLA in the United States[50], showing the extended lobe and plume emission (in yellow and green) with an extent of 80 kpc (16') and the inner lobes and jets (in red). The beam size of the image is 7.8" × 6.2"; the position angle is 86°. The inset shows the inner jet of M87 obtained by the EAVN precursor KaVA, on 2 March 2014[51]. The observation was made at 22 GHz. The restoring beam is 1.46 mas × 1.21 mas with a position angle of −3°, which is shown as a yellow oval in the bottom right corner of this inset. The naturally weighted image has 1σ image sensitivity of 0.4 mJy beam$^{-1}$. The contours represent 1.17 × (−1, 1, 2, $2^{3/2}$, $2^2$, $2^{5/2}$....) mJy beam$^{-1}$. The peak intensity is 1.37 Jy beam$^{-1}$. At the distance of M87, 16.7 Mpc, 1 mas angular size corresponds to a projected linear size of 0.08 pc or 140 Schwarzschild radii. The horizontal bar of an angular size of 5 mas in the KaVA/EAVN image represents a 1.3 light year. The KaVA image reveals that the inner jet is collimated at a few hundred gravitational radii from the central engine and extends to an angular distance of 20 mas. The full-operation EAVN expects to trace the jet out to a distance of 50 mas or further in the coming years with the enhanced sensitivity, enabling the velocity and spectral index evolution of the jet flow to be explored by virtue of its high-cadence multi-frequency capability. Credit: main image courtesy of NRAO/AUI and adapted from ref.[50], AAS/IOP; inset adapted from ref.[51], Oxford Univ. Press on behalf of the Astronomical Society of Japan.

**AGN and black hole studies.** The earliest VLBI observations were motivated to resolve compact structures in quasars and resulted in the discovery of bright, apparently superluminal jet knots ejected from the nuclear source[24,25]. Imaging of AGN structures is also a key science subject of the EAVN and its precursor KaVA. The KaVA AGN Large Program is intensively monitoring two key sources, M87 and Sgr A*. The jet stream launched from and collimated close to the SMBH in the centre of M87, a supergiant elliptical galaxy in the constellation Virgo[26], is observed on a biweekly basis at 22 and 43 GHz simultaneously. With the unprecedented high cadence and high-fidelity polarimetric VLBI images, the astronomers expect to get precise velocity field and spectral index evolution along and across the jet to constrain the magnetic field structure close to the jet launching site and related processes. At the border of the scatter-broadening regime of Sgr A* hosting a 4 ×$10^6$ M$_\odot$ SMBH in the centre of the Milky Way, KaVA provides the most sensitive images. These allow the radiation mechanism of the closest SMBH to be studied. The KVN by itself provides the unique astrometry of the frequency-dependent location of Sgr A*, and the KaVA AGN Large Program is producing the most sensitive milliarcseond-resolution images of Sgr A* at 43 GHz. Other AGN studies obtained with the EAVN subarrays are related to imaging of compact nuclear structures and studies of jet physics[27,28]. Even higher sensitivity with wider and agile frequency coverage of the EAVN enables astronomers to probe broader power range of AGNs, up to the highest redshifts in the early Universe[29].

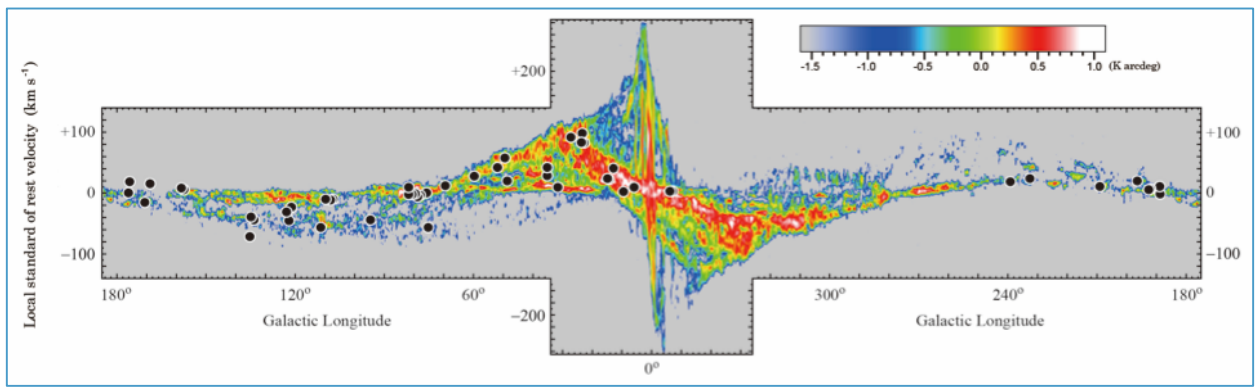

**Figure 5: Interstellar masers (harboured in star-forming regions) are used for determining the Galactic structure.** The coloured image shows the longitude–velocity diagram of the CO line emission[52]. The colour scale represents the logarithmic intensity from −1.5 to 1.0 (K arcdeg). The overlaid black dots mark the locations of 52 maser spots observed with VERA, VLBA and EVN by using the phase-referencing VLBI technique[32]. These data were employed to model the fundamental parameters of the Milky Way. LSR, local standard of rest. Credit: image adapted from ref.[52], AAS/IOP; maser spots adapted from ref.[32], Oxford Univ. Press on behalf of the Astronomical Society of Japan.

**Star formation and stellar evolution in the nearby Universe.** The discovery of astronomical masers in the microwave bands has encouraged VLBI studies, owing to their compact size and high brightness. Masers are often used as excellent probes to determine trigonometric distances to massive star-forming regions and evolved stars, to study the kinematics and magnetic fields of the interstellar and circumstellar environments around these regions [30]. High-accuracy radio astrometry of up to 100 water ($H_2O$) maser sources with VERA, VLBA and EVN has made it possible to determine the dynamical structure of the Milky Way [31,32]. VERA with the JVN and Nobeyama Radio Observatory (NRO) have made progress on methanol ($CH_3OH$) and silicon-monoxide (SiO) maser observations [33,34]. By linking the JVN to the Sheshan 25 m telescope, a large programme was conducted for measurement of internal proper motions of more than 30 methanol maser sources around young massive stars[35]. The KVN has operated a Key Science Program for simultaneous monitoring observations of circumstellar $H_2O$ and SiO masers at 22, 43, 86, and 129 GHz (http://radio.kasi.re.kr/kvn/ksp.php ). Two KaVA Large Programs (http://radio.kasi.re.kr/kava/large programs.php ) are on-going, aiming at yielding large samples of interstellar and circumstellar maser sources with intensive monitoring. Such large VLBI observation projects listed here focus on revealing dynamic behaviors of dense molecular gas outflows from massive young stars and pulsating dying stars in order to better understand the physics of interstellar and circumstellar masers and how high-mass stars grow and how stars end their lives. These projects, however, still target maser sources within a few kiloparsecs of the Solar System, while some of the sources at signposts of fast stellar evolution [36] are located at larger distances. Extragalactic masers (called megamasers) are brighter than Milky Way masers by up to eight orders of magnitude and 100–1,000 times more luminous than the Sun [37,38]. They have now been found for four molecules producing maser emission: hydroxyl (OH), $H_2O$, formaldehyde ($H_2CO$), and $CH_3OH$ associated with dense gas regions around AGNs or bursts of massive star formation. The EAVN is potentially a powerful facility to study such faint maser sources.

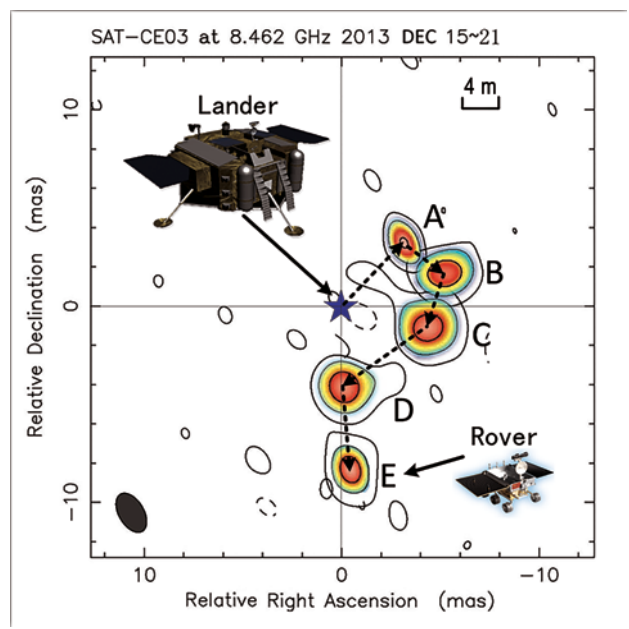

**Fig. 6: Motion of the Chang'E-3 rover with respect to the lander determined by phase-referencing VLBI, which produces an image at each parking site from A to E.** The colour scale represents the strength of beacon signals on the Yutu rover and increases from light blue to red. The lowest contour represents three times the image noise, and the contours increase in step of two. The blue star marks the position of the Chang'E-3 lander. The black dashed line denotes the rover's track on the Moon. The relative positional accuracy of the rover with respect to the lander is determined to be about 1 m (corresponding to an angular size of about 0.5 mas in the image). Credit: image adapted from ref. [39], Science China Press.

**Space exploration with VLBI.** Besides astrophysics and astrometry applications, VLBI has also been used for deep space exploration, offering orbit determination of spacecraft with unprecedented accuracy, and planetary science research, including atmosphere physics and geodynamics. A representative application is the successful tracking of Chinese lunar exploration satellites by the CVN in quasi-real time with the benefit of rapidly-developing digital recording and data acquisition, a software correlator, and real-time data transfer techniques. It is the first instance of VLBI navigation of a lunar exploration satellite. During the latest Chang'E-3 mission, the time delay between data acquisition at the telescopes and the delivery of the position measurement was less than one minute. The position of the Chang'E-3 lander on the surface of the Moon was determined with an accuracy of 100 m by using the VLBI monitoring data. A relative positional accuracy of 1 m was achieved for the distance between the Chang'E-3 lander and the rover (named Yutu) by using the phase-referencing technique [39] and analysing differential phase delays [40], the accuracy of which is ten times better than for the Apollo project. Figure 6 illustrates the moving track of the rover. The EAVN will broaden the collaborative potential in further deep space exploration missions.

**Supernovae, gamma-ray bursts and pulsars.** – Supernovae (SNe) and gamma-ray bursts (GRB) are the most energetic events in the Universe. A significant increase of the sample size of SNe and GRB is essential to understand their physical process [41]. Flexible coordination of the triggering is required to catch the earliest evolution of SNe and GRBs. VLBI parallax measurements of pulsars' distance have a typical accuracy of 100 µas per epoch, which converts to an uncertainty of only 2% for its distance determination [42]. A CVN−LBA project is underway to improve the distance accuracy of the pulsar J0437−4715, which will be used to set stringent constraints on the variation with time of Newton's gravitational constant $G$ [43]. Motivated by the discovery of magnetar and extended X-ray sources in the Galactic centre, high-frequency KVN observations have been conducted and a KaVA pulsar project is being discussed [44].

**Geodetic studies.** VLBI techniques found early applications in astrometry and geodesy [45], which have proven to be accurate and highly useful with three decades of application. The Asia-Oceania VLBI Group for Geodesy and Astrometry was established in 2014, coordinating geodetic observations with 17 telescopes. Eleven telescopes from the EAVN participate in the joint research, and two correlators at Shanghai Astronomical Observatory of China and Geospatial Information Authority of Japan undertake data processing tasks, a significant contribution to regional and global VLBI geodesy research.

## Future prospects of EAVN

The VLBI capabilities of the EAVN developed rapidly with the enhancement of correlation capability and building of new telescopes. Future extensions of the EAVN, including the addition of large telescopes and upgrades of existing equipment, will enable the EAVN to uniquely stand out among other VLBI arrays in terms of science and technological capabilities.

As mentioned earlier, the FAST 500 m telescope in southwest China, completed its engineering construction in September 2016[12]. Early science with FAST will focus on pulsar search, surveys of hydrogen lines and OH megamasers [46]. The addition of FAST to the EAVN will achieve milliarcsecond-scale resolution and an image noise of about 4.2 µJy/beam (1σ) at 2.3 GHz within one hour of integration time, 4.6 times better than the EAVN without FAST. Owing to the unprecedented imaging sensitivity, FAST−EAVN will play an extraordinary role in observing radio-quiet AGNs in the nearby Universe and radio-loud quasars in the early Universe, neutral hydrogen (H I) inflows and outflows in AGNs, detecting radio sources associated with tidal disruption events, probing the expansion of SNe, tracing the structural evolution of jets in X-ray binary and GRBs in long duration, precisely locating fast radio bursts and pulsars, discovering faint maser sources in the Milky Way and in distant galaxies, and searching for SMBH binaries and intermediate-mass black holes. In addition, construction of the 110 m QiTai Telescope has been initiated in Xinjiang Province in northwest China, with a wide frequency coverage from 0.15 to 115 GHz. The planned Thailand VLBI Network, consisting of four telescopes working at centimetre and millimetre wavelengths, will join the EAVN on completion. Located at lower latitudes, this array will facilitate joint observations between the EAVN and the LBA, forming a transcontinental Asia−Oceania VLBI network at centimetre and long millimetre wavelengths. In the north−south direction, a maximum baseline of > 10,000 km can be obtained with this network. Such growth of the EAVN will improve the capability of snapshot imaging, beneficial for studies of objects showing high temporal variability including transient objects.

At present, the EAVN is one of the few facilities that enable 3 mm VLBI observations. The construction of the fourth and fifth KVN telescopes is being considered to complete the full KVN project, which will enhance the sensitivity to extended emission structures. The Yonsei station of the KVN and the NRO 45 m telescope each plan to install a 230 GHz receiver, to significantly enhance the observational capability of the EAVN at 1.3 mm wavelength for participation in the EHT observations. The James Clerk Maxwell Telescope has been participating in the EHT collaboration at 1.3 mm since 2015 and is expected to observe with the EAVN. The Greenland telescope (12 m), the Solar Planetary Atmosphere Research Telescope (10 m) in Japan and the Seoul Radio Astronomy Observatory (6 m) in Korea may also join the EAVN, further enhancing the 1.3 mm capability. This synergy between the EAVN and the Global mm-VLBI Array, EHT and possibly the phased-up Atacama Large Millimeter/submillimeter Array would probe the strong-gravity regime around black holes, and directly image SMBH silhouettes in Sgr A* and M87 with superb angular resolution approaching the size of the black hole's event horizon.

Another approach to further increase the VLBI resolution is to enlarge the baseline lengths, by placing radio telescopes in space and enabling space VLBI together with ground-based stations, resulting in baselines spanning many times the Earth's diameter. Before the present VLBI collaboration of the EAVN, ground telescopes in Asia have extensively participated in the VLBI Space Observatory Programme (VSOP), which was led by astronomers in Japan and launched in 1997 [47]. These joint observation activities [48] formed a base for current VLBI activities in East Asia. VSOP also validated the key technologies of space VLBI and paved the way for future space VLBI missions. The Russia-led RadioAstron with a larger 10 m telescope was launched in 2011 and achieves a highest resolution of 8 μas at 22 GHz, which is a new milestone. Motivated by the success of VSOP and RadioAstron, proposals of next-generation space VLBI systems with higher sensitivity (larger space telescope), higher frequency (high resolution and less opacity) and denser $u$–$v$ coverage (more space elements) have been discussed [49]. A millimetre and submillimetre space VLBI array is free of atmospheric fluctuation of interferometer fringes on space–space baselines and onboard hydrogen-maser signals, and is essential for black hole astrophysics at sub-event horizon resolution, despite the technical challenges and tremendous cost. Another science-motivated direction is the development of high-sensitivity low-frequency space VLBI with a >30-m-level space radio telescope in operation at 0.3–5 GHz. It will lead to unique discoveries through synergistic observations with the large ground-based FAST 500 m telescope and the upcoming Square Kilometre Array (SKA); thus it is expected to open a new window for VLBI studies of weak radio AGNs, distant OH megamasers, and Galactic radio stars.

**Acknowledgement** T.A. thanks the grant supported by the Ministry of Science and Technology of China (2016YFE0100300), the Youth Innovation Promotion Association and FAST Fellowship of Chinese Academy of Sciences. H.I. thanks the grants supported by KAKENHI (16H02167) and the Korea Astronomy and Space Science Institute Commissioning Program. The authors are grateful to the engineering and scientific teams of the EAVN for test experiments. The authors thank Y. Hagiwara for his contribution in preparing the draft, W. Baan, D. Byun, R. Dodson, S. Frey, K. Fujisawa, K. Hada, M. Honma, D.R. Jiang, T.

Jung, M. Kino, S.-S. Lee, D. Li, P. Mohan, R.D. Nan, C.S. Oh, Z.H. Qian, M. Rioja, K. Shibata, F.W. Tong, K. Wajima, N. Wang, S.H. Ye, Y. Yonekura and Y.J. Yun for comments on the manuscript and for providing helpful information. B.W.S. is grateful for the support of the National Research Council of Science and Technology, Korea (EU-16-001).

**Author contributions** T.A. coordinated the writing of the paper. All authors have contributed to the EAVN commissioning and the preparation for this Perspective.